\documentclass{article}
\usepackage{lscape,epsfig}
\usepackage{graphicx}
\usepackage{amssymb}
\usepackage{amsmath}
\usepackage{natbib}
\usepackage{float}
\usepackage{subfloat}
\usepackage{rotating}
\textheight=22cm \DeclareSymbolFont{ppa}{OT1}{ppl}{m}{it}
\DeclareMathSymbol{\vv}{\mathalpha}{ppa}{'166}

minus 2.3mu \thickmuskip = 2.6mu plus 2mu minus 2.6mu

\begin{document}

\newcommand{\dd}{\,{\rm d}}
\newcommand{\ie}{{\it i.e.},\,}
\newcommand{\etal}{{\it $et$ $al$.\ }}
\newcommand{\eg}{{\it e.g.},\,}
\newcommand{\cf}{{\it cf.\ }}
\newcommand{\vs}{{\it vs.\ }}
\newcommand{\zdot}{\makebox[0pt][l]{.}}
\newcommand{\up}[1]{\ifmmode^{\rm #1}\else$^{\rm #1}$\fi}
\newcommand{\dn}[1]{\ifmmode_{\rm #1}\else$_{\rm #1}$\fi}
\newcommand{\upd}{\up{d}}
\newcommand{\uph}{\up{h}}
\newcommand{\upm}{\up{m}}
\newcommand{\ups}{\up{s}}
\newcommand{\arcd}{\ifmmode^{\circ}\else$^{\circ}$\fi}
\newcommand{\arcm}{\ifmmode{'}\else$'$\fi}
\newcommand{\arcs}{\ifmmode{''}\else$''$\fi}
\newcommand{\MS}{{\rm M}\ifmmode_{\odot}\else$_{\odot}$\fi}
\newcommand{\RS}{{\rm R}\ifmmode_{\odot}\else$_{\odot}$\fi}
\newcommand{\LS}{{\rm L}\ifmmode_{\odot}\else$_{\odot}$\fi}

\newcommand{\Abstract}[2]{{\footnotesize\begin{center}ABSTRACT\end{center}
\vspace{1mm}\par#1\par \noindent {~}{\it #2}}}

\newcommand{\TabCap}[2]{\begin{center}\parbox[t]{#1}{\begin{center}
  \small {\spaceskip 2pt plus 1pt minus 1pt T a b l e}
  \refstepcounter{table}\thetable \\[2mm]
  \footnotesize #2 \end{center}}\end{center}}

\newcommand{\TableSep}[2]{\begin{table}[p]\vspace{#1}
\TabCap{#2}\end{table}}

\newcommand{\FigCap}[1]{\footnotesize\par\noindent Fig.\  %
  \refstepcounter{figure}\thefigure. #1\par}

\newcommand{\TableFont}{\footnotesize}
\newcommand{\TableFontIt}{\ttit}
\newcommand{\SetTableFont}[1]{\renewcommand{\TableFont}{#1}}
\newcommand{\MakeTable}[4]{\begin{table}[htb]\TabCap{#2}{#3}
  \begin{center} \TableFont \begin{tabular}{#1} #4
  \end{tabular}\end{center}\end{table}}

\newcommand{\MakeTableSep}[4]{\begin{table}[p]\TabCap{#2}{#3}
  \begin{center} \TableFont \begin{tabular}{#1} #4
  \end{tabular}\end{center}\end{table}}

\newenvironment{references}%
{ \footnotesize \frenchspacing
\renewcommand{\thesection}{}
\renewcommand{\in}{{\rm in }}
\renewcommand{\AA}{Astron.\ Astrophys.}
\newcommand{\AAS}{Astron.~Astrophys.~Suppl.~Ser.}
\newcommand{\ApJ}{Astrophys.\ J.}
\newcommand{\ApJS}{Astrophys.\ J.~Suppl.~Ser.}
\newcommand{\ApJL}{Astrophys.\ J.~Letters}
\newcommand{\AJ}{Astron.\ J.}
\newcommand{\IBVS}{IBVS}
\newcommand{\PASP}{P.A.S.P.}
\newcommand{\Acta}{Acta Astron.}
\newcommand{\MNRAS}{MNRAS}
\renewcommand{\and}{{\rm and }}
\section{{\rm REFERENCES}}
\sloppy \hyphenpenalty10000
\begin{list}{}{\leftmargin1cm\listparindent-1cm
\itemindent\listparindent\parsep0pt\itemsep0pt}}%
{\end{list}\vspace{2mm}}

\def\TYLDA{~}
\newlength{\DW}
\settowidth{\DW}{0}
\newcommand{\dw}{\hspace{\DW}}

\newcommand{\refitem}[5]{\item[]{#1} #2%
\def\REFARG{#3}\ifx\REFARG\TYLDA\else, {\it#3}\fi
\def\REFARG{#4}\ifx\REFARG\TYLDA\else, {\bf#4}\fi
\def\REFARG{#5}\ifx\REFARG\TYLDA\else, {#5}\fi.}

\newcommand{\Section}[1]{\section{#1}}
\newcommand{\Subsection}[1]{\subsection{#1}}
\newcommand{\Acknow}[1]{\par\vspace{5mm}{\bf Acknowledgements.} #1}
\pagestyle{myheadings}

\newfont{\bb}{ptmbi8t at 12pt}
\newcommand{\xrule}{\rule{0pt}{2.5ex}}
\newcommand{\xxrule}{\rule[-1.8ex]{0pt}{4.5ex}}
\def\thefootnote{\fnsymbol{footnote}}

\begin{center}
{\Large\bf
 The Clusters AgeS Experiment (CASE). \\
 Variable stars in the field of the open cluster NGC~6253\footnote{Based
 on data obtained at Las Campanas Observatory.}}
 \vskip1cm
  {\large
      ~~J.~~K~a~l~u~z~n~y$^1$,
      ~~M.~~R~o~z~y~c~z~k~a$^1$,
      ~~W.~~P~y~c~h$^1$
      and ~~I.~B.~~T~h~o~m~p~s~o~n$^2$,
   }
  \vskip3mm
$^1$Nicolaus Copernicus Astronomical Center, ul. Bartycka 18, 00-716 Warsaw, Poland\\
     e-mail: (jka, mnr, wp)@camk.edu.pl\\
  $^2$The Observatories of the Carnegie Institution of Washington, 813 Santa Barbara
      Street, Pasadena, CA 91101, USA\\
     e-mail: ian@obs.carnegiescience.edu\\
\end{center}

\vspace*{7pt}
\Abstract
{The field of the metal-rich open cluster NGC~6253 has been surveyed in a search for 
variable stars. A total of 25 new variables were detected, 14 of which are bright stars with 
$13<V<15$ mag. This domain was not covered in an earlier work by de Marchi et al. (2010).
Four variables, including three short-period eclipsing binaries, are 
candidate blue straggler stars. Two new detached eclipsing binaries at the  turnoff of the 
cluster and another one on the subgiant branch were identified. These three systems deserve
a detailed follow-up study aimed at a determination of the age and distance of NGC~6253. 
New photometry for 132 stars from the sample of de Marchi et al. (2010) is provided.
}
{open clusters: individual (NGC 6253) -- stars:variables --
blue stragglers -- binaries: eclipsing)
}

\Section{Introduction} 
\label{sec:intro}
NGC~6253 ($l=335.4$, $b=-6.3$) is an old open cluster projected against the 
central part of the Galactic disc. A first study based on $UBVRI$ CCD photometry 
was presented by Bragaglia et al. (1997). An important result of this
seminal paper was  strong evidence that the metallicity of the cluster 
is about twice the solar value. Subsequent photometric studies were published 
by Piatti et al. (1998; $BVI$), Sagar et al. (2001; $UBVRI$), 
Twarog et al. (2003; $uvbyH\beta$), and Montalto et al. (2009; $BVRIJHK$). The 
latter paper includes proper motion (PM) data and membership probabilities for brighter 
stars in the cluster field. Montalto et al. (2009) also discuss the age of 
NGC~6253, quoting 3.5 Gyr as the most likely value. 

Based on spectra of four red-clump giants, Carretta et al. (2007) determined the 
metallicity of NGC~6253 to be ${\rm  [Fe/H]}=+0.46\pm 0.03\pm 0.08$ (rms + systematic 
error). The same value of ${\rm  [Fe/H]}=+0.46^{+0.02}_{-0.03}$ was derived from 
high resolution spectra of 65 stars by Anthony-Twarog et al. (2010). Spectra of
seven stars obtained by Sestito et al. (2007) yielded ${\rm  [Fe/H]} =+0.37\pm 
+0.07$ - a value slightly lower, but consistent within error limits with the 
previous two measurements. An even lower value was more recently reported by Montalto et al. 
(2012), who measured ${\rm  [Fe/H]}=+0.26\pm 0.11$ and 
${\rm [Fe/H]}=+0.19\pm0.13$ from VLT/UVES high-resolution spectra of a 
main-sequence star and two red clump stars, respectively.

The first and so far the only search for variable stars in NGC~6253 
field was conducted by De Marchi et al. (2010). They used ESO-MPI 2.2-m 
telescope plus wide field CCD WFI camera to monitor a $32\times 32$ arcmin
region centered on the cluster. A total of 45.3 hours of observations 
in $R$-filter were collected over  10 consecutive nights in 2004.
Light curves were extracted for about 250 000 stars of which 595 were
classified as variables. The sample of variables located within 8 arcmin
from the cluster centre and considered possible members includes
13 contact binares, 9 detached or semi-detached systems, 16 ``rotational'' 
variables, 16 long-period objects and one dwarf nova. In our opinion none 
of the detached binaries listed by De Marchi et al. (2010) deserves a 
detailed follow-up study.

We obtained an independent photometry of NGC~6253  within the CASE project
(Kaluzny et al. 2005) with the aim to detect detached 
eclipsing binaries which could be used for the determination of age and distance 
of the cluster, as proposed by (Paczy\'nski 1997). Before we reduced our data, 
Montalto et al. (2011) discovered that their star \#45368 -- a proper-motion 
and radial-velocity member of NGC~6253 located at the turnoff of the cluster --
is an eclipsing binary. This star was not included among variable stars 
of De Marchi et al. (2010). We decided to focus 
on this object in order to obtain complete light and velocity curves (which
will be analysed in a separate contribution). Here we present light 
curves of 25 new variable stars detected in the cluster field. We have also identified 
132 stars from the list of variables compiled by De Marchi et al. (2010). Their 
light curves and periods are available from the CASE 
archive\footnote {http://case.camk.edu.pl/}.  

While searching the literature for reliable photometric data that would 
enable a transformation
from our instrumental system to the standard one, we found large discrepancies 
between published sets of $BV$ photometry of NGC~6253. We discuss these discrepancies 
and argue in favor of a transformation to the standard $BV$ system based on the 
photometry of Sagar et al. (2001). 

\section{Observations}
\label{sec:obs}
The data were collected at Las Campanas Observatory on 78 nights between 
August 2007 and October 2010. We used the 1.0-m Swope telescope and the SITE3 
CCD camera which has a scale of 0.445 arcsec/pixel. Time-series observations 
were conducted with $BV$ filters. We obtained a total of 1016 useful images in 
$V$ and 275 in $B$. The median exposure time was 90 s for $V$ and 150 s 
for $B$. The actual exposure time varied according to observing 
conditions,  longer for nights with clouds or poor seeing. The median 
seeing was equal to 1.5 and 1.6 arcsec for $V$ and $B$ frames, respectively. 
Images were taken with the camera subrastered to $2048\times 2150$ pixels,
resulting in a field of view of $14.8\times  15.6$ arcmin. 
The monitored region was centered at $RA(2000)=254.8053$~deg
and $Dec(2000)=-52.7200$ ~deg. These coordinates  correspond 
approximately to the center of NGC6253. Our field is included entirely 
within the region surveyed by De Marchi et al. (2010), covering 21\% of its
area. The angular diameter of NGC 6253 does not exceed 16 arcmin (Bragaglia 
et al. 1997), what implies that we surveyed nearly the whole area of the 
cluster.

\Section{Photometric Reductions}
\label{sec:photometry}
The full field of $2048\times 2150$ pixels was divided into a mosaic of 
$5\times 5$ overlapping subfields. These subfields were analysed 
independently, thereby reducing the effects 
of PSF variations across the frame. 
Each subfield had a size of $570\times 590$ pixels
and the overlap between fields amounted to 160 pixels in X and Y.
The time-series photometry was derived with 
an image subtraction technique using the DIAPL code\footnote{Freely accessible at 
http://users.camk.edu.pl/pych/DIAPL/index.html.}. Template images were obtained by 
averaging several frames taken with good sky conditions and good seeing. For the 
templates, profile photometry with Daophot and Allstar codes (Stetson 1987) was 
extracted. Aperture corrections necessary for the proper transformation of light 
curves from differential counts to instrumental magnitudes were derived for each 
subfield separately using the Daogrow code (Stetson 1990). 

\begin{figure}[htb]
\centerline{\includegraphics[width=0.95\textwidth, bb= 25 171 564 694,clip]{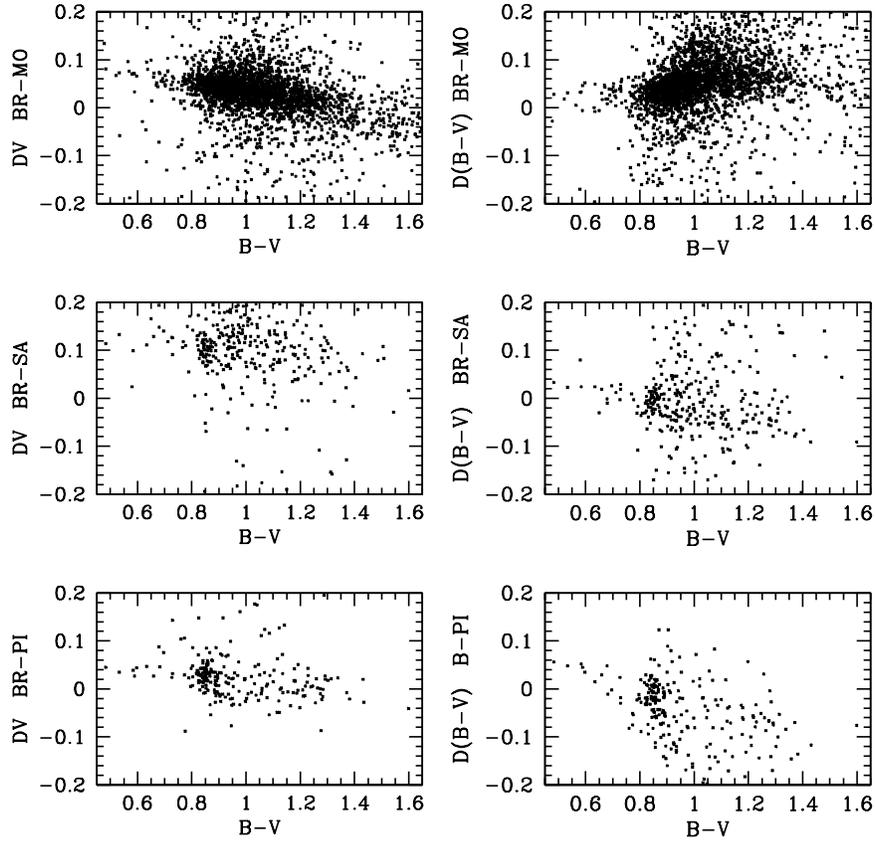}}
\caption{\small
Differences between NGC 6253 photometries of Montalto et al. (2010), Sagar et al. (2001) 
and Piatti et al. (1998) vith respect to  the photometry of Bragaglia et al. (1997) are 
shown in top, middle and bottom rows, respectively, as a function of the $B-V$ color. 
The left panel of each row shows differences in $V$, while the right one -- differences 
in $B-V$. All differences are clearly color-dependent.
 \label{fig:phot_comp}
}
\end{figure}

We were planning to transform the instrumental system to the standard one using 
results from existing photometric studies of NGC~6253. Unfortunately, as noted 
already by Anthony-Twarog et al. (2010), the four available sets of $BV$ 
photometry mentioned in Sect. \ref{sec:intro} show significant discrepancies. 
We compared these  using data from the WEBDA 
base\footnote{http://www.univie.ac.at/webda/navigation.html} to find 
that the differences are not only large, but also color-dependent 
(Fig.~\ref{fig:phot_comp}). A detailed comparison indicates
that no two of the four data sets agreed with each other. 
Photometry of Piatti et al. (1998) was calibrated based on 
12 standard stars from Landolt (1992). Bragaglia et al. (1997) 
used data for 18 stars from four Landolt fields. Montalto et al. (2009)
observed one Landolt field, but they based their calibration on Stetson's 
photometry\footnote{http://cadcwww.dao.nrc.ca/standards/}. 
The calibration of Sagar et al. (2001) is based on 40 
standard stars from  E and F regions (Menzies et al. 1989).

Within the CASE project we have always taken a great care to accurately 
transform our photometry from the instrumental system to the standard one. 
Specifically, we have used equations 
\begin{eqnarray}
v&=&a_1+V+a_2\times(B-V)+X\times k_{v}\\
b&=&a_3+V+a_4\times(B-V)+X\times k_{b}\\
b-v&=&a_5+a_6\times(B-V)+X\times k_{bv},
\end{eqnarray} 
where $v$ and $b$ are instrumental magnitudes, $X$ is the air mass, and all 
coefficients are to be found from observations of photometric standards. 
During most of the CASE observing
seasons the same CCD camera with the same set of filters was used, and in 
several photometric nights observations of Landolt standards were taken. We found 
that over the years the color terms in (1)-(3) remained constant. Facing serious 
discrepancies in the published photometry of NGC~6253, we decided to adopt 
as secondary standards the data set which would lead to a transformation with color 
terms most consistent with ours. This was done by substituting in (1), (2) and (3)
our instrumental $bv$ magnitudes together with the corresponding published  
magnitudes and calculating the coefficients. As it turned out, only the 
photometry of Sagar et al. (2001) reproduced $a_2$, $a_4$ and $a_6$ in a consistent 
fashion, yielding  
\begin{eqnarray}
v&=&V-1.679(7)-0.008(7) (B-V)\\ 
b&=&B-2.090(9)-0.046(9) (B-V)\\
(b-v)&=&-0.413(7)+0.964(7) (B-V),
\end{eqnarray}  
with the terms $Xk$ from equations (1)-(3) included in additive constants.
The transformations (4)-(6) are based on 122 stars with $11.6<V<18$ and
$0.27<B-V<1.79$. The $rms$ values of residuals for these stars
amount to 0.019, 0.022 and 0.018 mag for $V$, $B$ and $B-V$, respectively.  
We therefore transformed all our instrumental data to the system of Sagar et al. 
(2001). We are unable to verify the correctness of zero points of this 
transformation, but we hope that it is free from systematic color-dependent errors. 
However, it is clear that new observations are needed to clarify the calibration 
problems discussed above.

\begin{figure}[htb]
\centerline{\includegraphics[width=0.95\textwidth, bb= 21 236 495 498,clip]{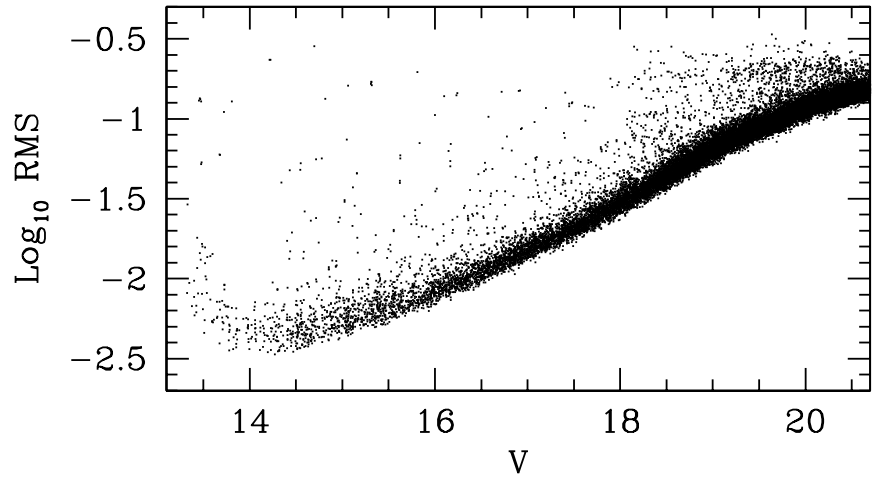}}
\caption{\small
The accuracy of the photometry of the observed field. {\it RMS} values of individual 
measurements in the $V$-band are plotted vs. the average $V$-magnitude.
 \label{fig:phot_acc}}
\end{figure}

\begin{figure}[htb]
\centerline{\includegraphics[width=0.95\textwidth, bb= 20 169 571 690,clip]{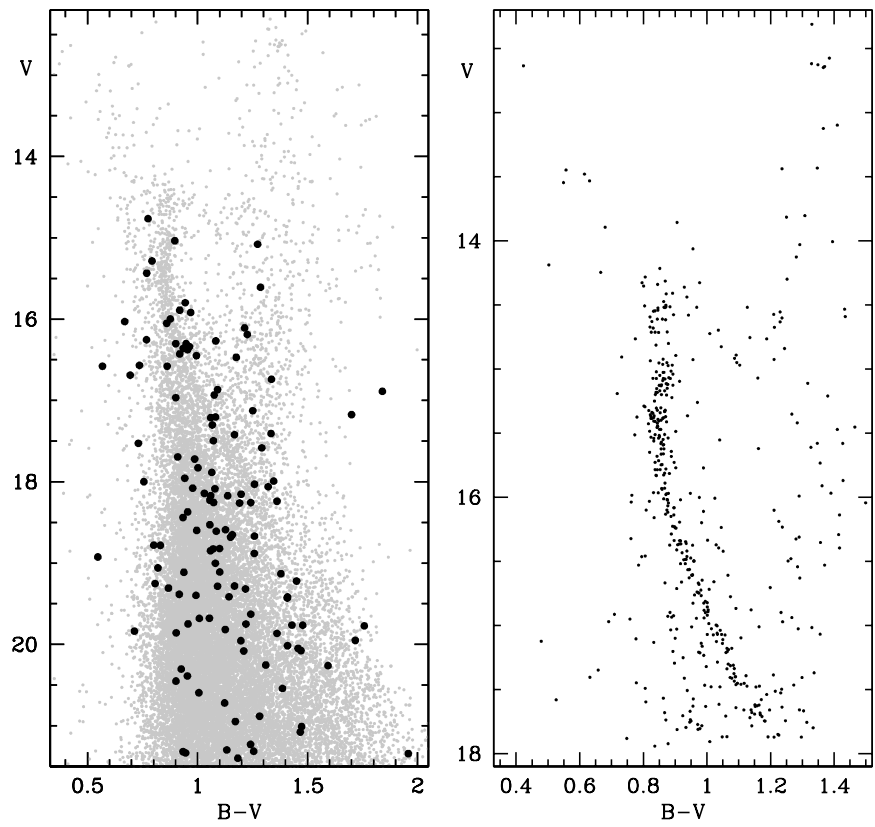}}
\caption{\small
Color-magnitude diagram of the observed field. Left: all stars (variable stars discovered 
by De Marchi 
et al. (2010) are marked with thick black dots); right: proper motion members
of NGC 6253 (note the change of the vertical scale). \label{fig:cmd}}
\end{figure}

As shown in Fig.~\ref{fig:phot_acc}, the accuracy 
of our photometry decreases from 3 mmag at $V$ = 14 mag to 25 mmag at $V$ = 
18 mag and 100 mmag at $V$~=~20~mag. Fig.~\ref{fig:cmd} shows the color-magnitude 
diagram (CMD) of the observed field. The right panel only includes objects which 
are PM-members of the cluster according to Montalto et al. (2009), with the 
membership probability limit decreasing linearly from 90\% at $V=12.5$ to 50\% 
at $V=18.0$. The PM data were only available for $\sim$37\% of the observed field.   

The astrometric solution for the $V$-band reference image  was determined from the  
positions of 1400 UCAC4 stars (Zacharias et al. 2013). The
average residuals in RA and DEC between the cataloged and recovered 
coordinates amount to $0.000\pm 0.077$ and $0.000\pm 0.125$ arcsec,
respectively.

\Section{Variable Stars}
 \label{sec:variables}
We derived $V$-band light curves of 21871 stars with $V<20.5$, each  
of these containing at least 500 measurements with a magnitude error $\leq0.1$ mag. 
We began by using our equatorial coordinates to identify variables already reported 
by de Marchi et al. (2010). In the case of doubtful identification, light curves of the 
best matching objects were examined to see which is the actual variable. 
We identified 132 objects in this way. The light curves of the remaining stars from our 
sample were examined for periodic variability with the AoV and AOVTRANS algorithms 
(Schwarzenberg-Czerny 1996, Schwarzenberg-Czerny \& Beaulieu 2006) as implemented in 
the TATRY code\footnote{http://users.camk.edu.pl/alex/\#software}. Twenty-six variables 
not listed in de Marchi et al. (2010) were 
identified. One of these is the eclipsing binary reported as star 
\#45368 by Montalto et al. (2011). 

The basic properties of the newly discovered variables are listed in Table~1 together
with their equatorial coordinates. Magnitudes at maximum at minimum light are, 
respectively, medians from the brightest 10\% and the dimmest 5\% of lightcurve points. 
Periods and period errors have been derived with the help of the TATRY code. 
The photometry of the new variables can be downloaded from
the CASE archive at http://case.camk.edu.pl. The archive
also contains the reference frame of the surveyed field taken
in $V$-band, and rectangular coordinates of all objects for
which the $V$-magnitude was obtained. This enables interested
observers to make tailored finding charts for the new variables.

Phased light curves of 22 
objects are presented in Figs.~\ref{fig:lcp1} and \ref{fig:lcp2}. Periods of the 
remaining four variables could not have ben determined, and the corresponding light curves
are shown in Fig.~\ref{fig:lct} as functions of time. The CMD of the cluster with  
the variables marked is shown in Fig.~\ref{fig:cmd_var}, divided into two panels 
to facilitate the identification. For clarity, only stars classified by Montalto et al. (2009)
as members of NGC 6253 are plotted 
as the background, but it is evident that the CMD is still contaminated by numerous field 
objects. This is due to a lack of clear separation of cluster and field stars on the 
vector-point diagram (Montalto et al. 2009). A new PM survey with longer time-baseline 
would improve this situation. 
 
Most of the new variables are bright stars with $V\lesssim15$. This is a natural consequence 
of the fact that the survey of de Marchi et al. (2010) was limited to stars fainter 
than $V\approx 15$ (see Fig. 3). Our sample includes four possible blue stragglers (BS) of which two, 
V13 and V14, are PM members of the cluster. The membership status of the remaining two is 
unknown. Three BSs, V11, V13 and V14 are eclipsing binaries. All 
three show shallow eclipses, indicating low mass ratios. Because the secondary eclipse of 
V11 is total, the mass ratio of this binary can be uniquely 
determined, and using Phoebe package (Pr\v sa \& Zwitter 2005) 
we obtained $q=0.16$ from the analysis of the $V$-band light curve.
The brightest of the variable BS candidates is the multiperiodic $\delta$~Sct-type pulsator 
V00 with a dominant periodicity of 0.030~d. The object is located about 0.5~mag above 
the extension of the unevolved main-sequence of the cluster. If it really belongs to 
NGC~6253, then the cluster apparent distance modulus of $11.8\pm 0.3$ (Twarog et al. 2003) implies
an absolute magnitude $M_{\rm V}=1.1$ mag for V00,  well within the limits of the absolute 
magnitudes of $\delta$~Sct stars (Handler 2005).

The contact binary V5 is located at the turnoff of the cluster together with detached 
binaries V15 and V16. Another detached system, V23, resides in the vicinity of the turnoff.
V15, V16 and V23 are PM-members of NGC~6253. As mentioned above, an analysis of V15 will be presented in a separate 
contribution. For V16 we partly covered five eclipses. This turned out to be insufficient 
for a unique determination of the period, so that the 4.58 d quoted in Table 2 should be regarded 
as a preliminary value. For V23 only one eclipse was observed, with a depth of 
about 0.12 mag.

By far the most interesting of our eclipsing binaries is V17, located on the subgiant 
branch of the cluster. The secondary eclipse is shallow with $\delta V\approx 0.15$~mag, 
but this does not exclude the possibility that the system is of SB2 type. The deep primary 
eclipse ($\delta V\approx 1$ mag) and different temperatures of the components should 
result in a unique solution of the light curves. More observations are needed 
to refine the shape of the secondary eclipse, but these should be relatively easy given the 
known ephemeris. Two other variables located on the subgiant branch are V19 and V20, 
the latter being a PM member of the cluster. Unstable sine-like light curves with 
low amplitudes indicate that these stars are spotted variables.
Several variables located off the main sequence and subgiant branch of the cluster are likely 
field objects. This grup includes the RRc star V07. As it was noted by the 
referee, the light curve of this star shows in fact two closely spaced 
periodicities bracketing the period listed in Table 1.

As a by-product of the present survey we obtained light curves of 132 variables from the 
list compiled by De Marchi et al. (2010) and analyzed these in  search for periodicity.
For most of them we confirmed the periods obtained by de Marchi et al. (2010), however in 
some cases alternative periods were found or no periodicity was detected. These results 
are available from the CASE archive at http://case.camk.edu.pl/.  


\Section{Summary}
\label{sec:sum}
We have conducted an extensive photometric survey of the metal rich open cluster NGC~6253 
in a search for variable stars. Twenty five new variables were detected. The sample 
includes 15 stars with $13<V<15$, thus extending an earlier investigation by De Marchi 
et al. (2010) to brighter objects. Three eclipsing binaries and one pulsating star were 
found in the blue-straggler region. Three new detached eclipsing binaries were discovered 
in the turnoff region, and another one on the subgiant branch. These systems are  
promising candidates for spectroscopic and photometric follow-up to determine  
age and distance modulus of the cluster. Light curves and revised periods were obtained 
for 132 variables from the sample of De Marchi et al. (2010).
 These are available from the CASE electronic archive together 
with our photometric data for NGC~6253. 

\Acknow
{
JK, MR and WP were partly supported by the grant DEC-2012/05/B/ST9/03931
from the Polish National Science Center. We thank the referee
for the constructive and helpful report. 
}

\begin{sidewaystable}
\begin{center}
\caption{
Basic data of NGC~6253 variables identified within the present survey
}
\begin{tabular}{|c|c|c|c|r|c|l|l|l|}
\hline
ID &  RA  & DEC & MP$^a$ &$V_{\rm max}$& ~$<B>$ &$DV$ &Period[d] & Type$^b$\\
    &      [deg]  & [deg] &  \%     &             &  $-<V>$    &     &          &         \\
\hline
00 & 254.78927 & -52.83545 & - &  12.860& 0.372& 0.07 &0.0304328(1) &$\delta$ Sct, BS \\
01 & 254.63200 & -52.77554 & - &  17.935& 0.947& 0.14 &0.2856667(1) & EW\\
02 & 254.87423 & -52.68190 & - &  19.442& 1.098& 0.51 &0.3201930(1) & EW\\
03 & 254.63978 & -52.61062 & - &  18.914& 1.168& 0.36 &0.3414981(1) & EW\\
04 & 254.64443 & -52.75881 & - &  20.051& 0.986& 0.70 &0.3440106(4) & EW\\
05 & 254.87089 & -52.66645 & - &  14.450& 0.843& 0.07 &0.35801236(1) & EW\\
06 & 254.79032 & -52.83517 & - &  16.717& 0.853& 0.20 & 0.39768833(1) & EW\\
07 & 254.63988 & -52.82878 & - &  16.587& 0.616& 0.43 &0.41130522(1) & RRc\\
08 & 254.67979 & -52.66643 & 0 &  14.619& 0.788& 0.035 &0.47518561(8) & Sp, Ell?\\
09 & 254.63440 & -52.61026 & - &  15.648& 0.626& 0.023 &0.47911411(1) & EW\\
10 & 254.67110 & -52.82267 & - &  17.999& 0.895& 0.31 &0.53285385(2) & EW\\
11 & 254.88926 & -52.65563 & - &  13.716& 0.664& 0.38 &0.57914361(1) & EW, BS\\
12 & 254.83146 & -52.64999 & - &  15.104& 0.878& 0.12 &0.6542998(6) & Sp\\
13 & 254.81845 & -52.71808 & 96&  13.653& 0.665& 0.36 &0.88644025(1) & EB, BS\\
14 & 254.75385 & -52.71431 & 95&  13.472& 0.612& 0.18 &1.15518598(1) & EA, BS\\
15 & 254.82151 & -52.71217 & 95&  14.709& 0.839& 0.27 &2.572391(2) & EA\\
16 & 254.77604 & -52.70025 & 97&  14.544& 0.832& 0.20 &4.580597(2) &  EA \\
17 & 254.62174 & -52.61255 & - &  14.669& 0.998& 1.01 &5.11162(3) & EA\\
18 & 254.67926 & -52.65474 & 56&  15.106& 1.822& 0.18 &25.501(3) & Sp\\
19 & 254.63715 & -52.60220 & - &  14.643& 1.104& 0.10 &27.4762(7) & Sp\\
20 & 254.75563 & -52.70552 & 95&  13.649& 1.292& 0.17 &28.2517(3) & Sp\\
21 & 254.85541 & -52.79515 & 91&  13.363& 1.754& 0.43 &39.491(1) & Sp\\
22 & 254.96731 & -52.83655 & - &  18.179& 2.623& $>2.0$ &- & P?\\
23 & 254.79213 & -52.72136 & 94&  15.198& 0.869& 0.13 &- & EA\\
24 & 254.68618 & -52.66308 & 98&  14.101& 1.679& $>0.73$ &- & P\\
25 & 254.97043 & -52.67775 & - &  14.825& 1.785& $>0.64$ &- & P\\
\hline
\end{tabular}
\end{center}
{\footnotesize $^a$ membership probability from 
Montalto et al. (2009); $^b$ type: EW - contact binary, EB - close eclipsing binary,
EA - detached eclipsing binary, BS - blue straggler candidate, Sp - spotted, Ell - elliptical,
 RRc - RR Lyr c-type, P - pulsating (long period)}
\end{sidewaystable}

\clearpage

\begin{subfigures}
\begin{figure}[htb]
\centerline{\includegraphics[width=0.95\textwidth, bb= 39 151 518 706, clip]{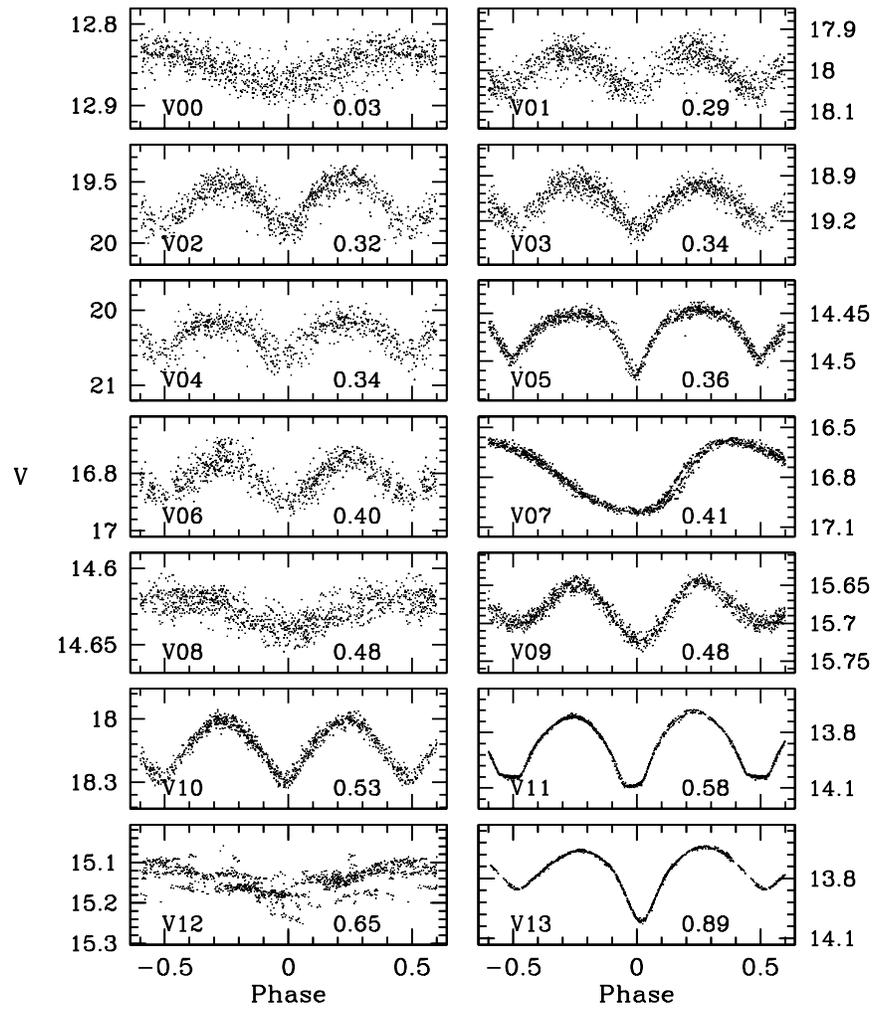}}
\caption{\small \label{fig:lcp1}
Phased $V$-light curves of new variables from the observed field (continued in 
Fig. \ref{fig:lcp2}).} 
\end{figure}

\begin{figure}[htb]
\centerline{\includegraphics[width=0.95\textwidth, bb= 39 377 518 706, clip]{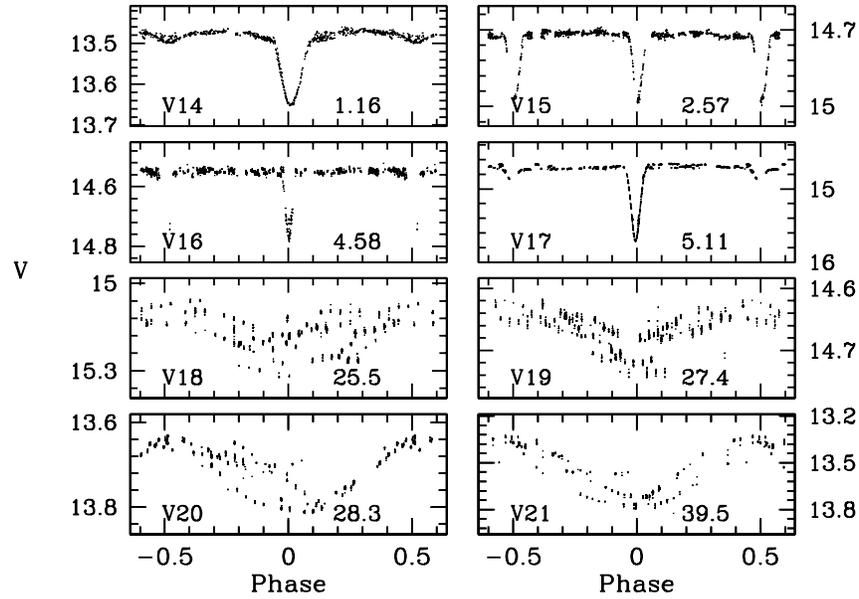}}
\caption{\small \label{fig:lcp2}
Phased $V$-light curves of new variables from the observed field (continuation of 
Fig. \ref{fig:lcp1}). Included is the detached eclipsing binary V15 discovered by Montalto 
et al. (2011; their star \#45368), which we analyse in detail in a separate contribution. } 
\end{figure}
\end{subfigures}

\begin{figure}[htb]
\centerline{\includegraphics[width=0.95\textwidth, bb= 49 376 477 706, clip]{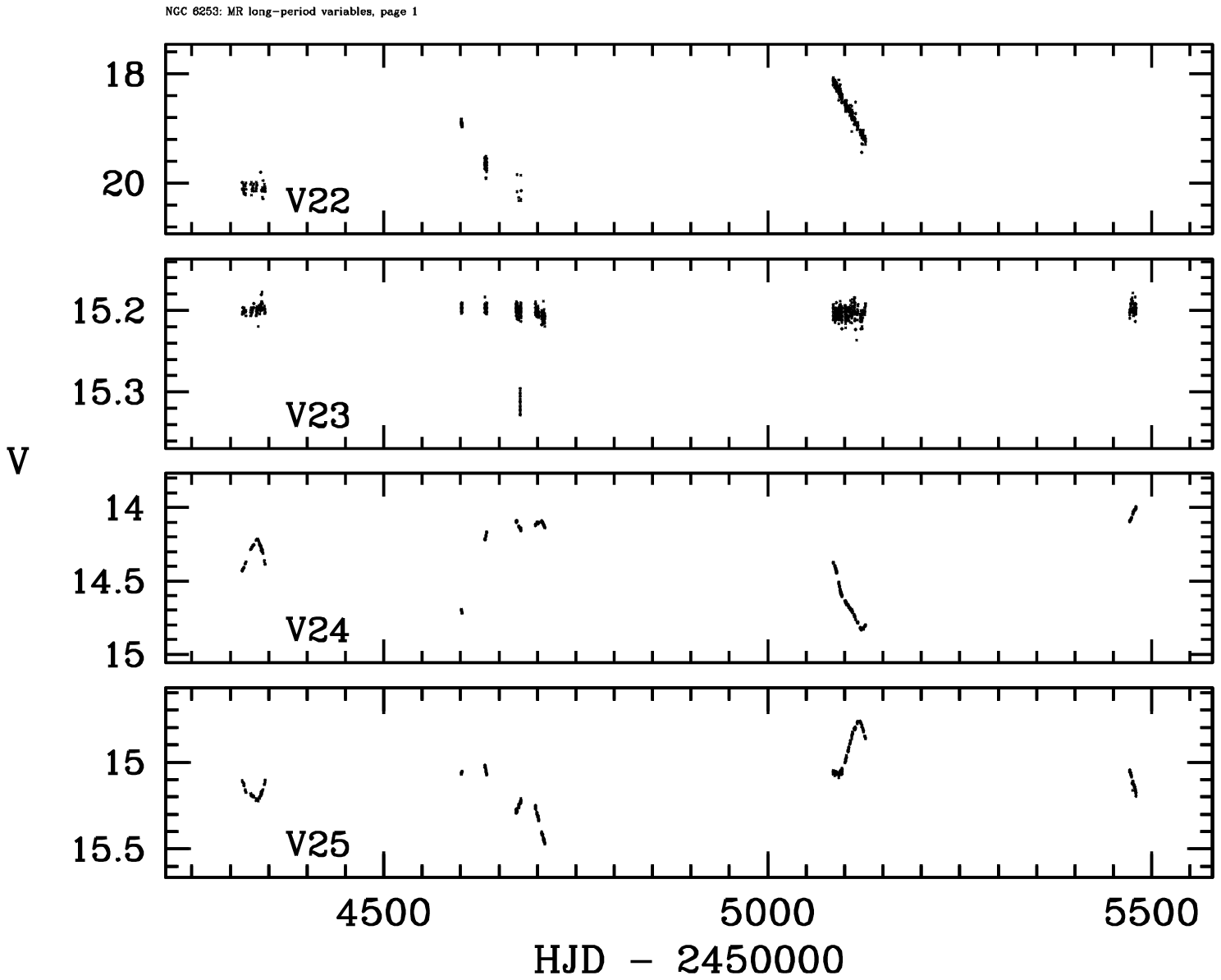}}
\caption{\small \label{fig:lct}
$V$-light curves of new long-period variables from the observed field.} 
\end{figure}

\begin{figure}[htb]
\centerline{\includegraphics[width=0.95\textwidth, bb= 51 48 564 742, clip]{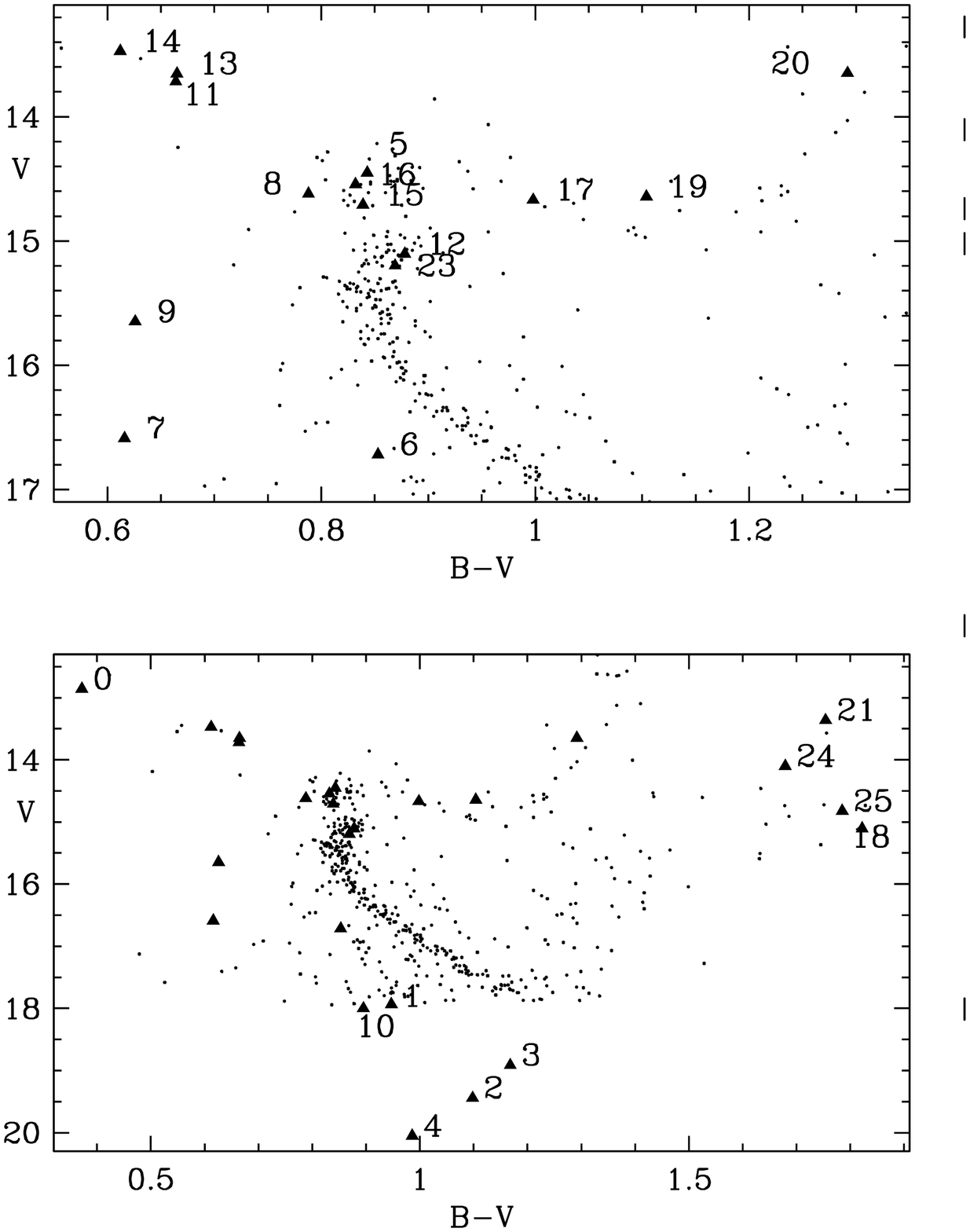}}
\caption{\small
Color-magnitude diagram for proper motion members of NGC~6252.
Positions of stars with light curves shown in Figs. \ref{fig:lcp1} and \ref{fig:lcp2} 
are marked with triangles and labeled. The red pulsating star V22 with $B-V=2.6 $ is not 
shown. Magnitudes at maksimum light and average colors from Table 1 are 
plotted. 
\label{fig:cmd_var}}
\end{figure}

\end{document}